\documentclass[alpha-refs]{wiley-article}

\usepackage{color}
\usepackage{ulem}

\usepackage{siunitx}
\usepackage{hhline}
\usepackage{multicol}
\usepackage{url,booktabs,multirow}
\usepackage{caption}
\usepackage{subcaption}
\usepackage{natbib}
\usepackage{color}
\usepackage{enumitem}
\usepackage{lineno,hyperref,amsmath}

\papertype{Original Article}

\title{Intercensal updating using structure-preserving methods and satellite imagery}

\author[1]{Till Koebe}
\author[1]{Alejandra Arias-Salazar}
\author[1]{Natalia Rojas-Perilla}
\author[1]{Timo Schmid}

\affil[1]{Department of Economics, Freie Universität Berlin, Berlin, Germany}

\corraddress{Till Koebe, Department of Economics, Freie Universität Berlin, Garystrasse 21, 14195 Berlin, Germany}
\corremail{till.koebe@fu-berlin.de}

\fundinginfo{The authors did not receive any specific funding for this work.}

\runningauthor{Koebe et al.}

\begin{document}

\maketitle

\begin{abstract}
Censuses are fundamental building blocks of most modern-day societies, yet collected every ten years at best. We propose an extension of the widely popular census updating technique \textit{Structure Preserving Estimation} by incorporating auxiliary information in order to take ongoing subnational population shifts into account. We apply our method by incorporating satellite imagery as additional source to derive annual small-area updates of multidimensional poverty indicators from 2013 to 2020 for a population at risk: female-headed households in Senegal. We evaluate the performance of our proposal using data from two different census periods.

\keywords{Multidimensional poverty, Official statistics, Small area estimation, SPREE} 
\end{abstract}

\section{Introduction}\label{section-introduction}

The estimation and monitoring of key statistical indicators are essential for the definition and adequate execution of public policies. How many young women without educational attainment currently live in district A? Answers to questions like this inform local policy-making on a daily basis. On a global scale, the 2030 Agenda for Sustainable Development highlights the need for information to be disaggregated not only geographically, but also along other dimensions such as sex, age, and employment status to identify and reach the most vulnerable in a society (\cite{UnitedNationsGeneralAssembly2015ResDevelopment}). Statistical systems around the world increasingly face growing demand for statistical information with specific quality characteristics such as disaggregation, periodicity and relevance in order to guide policy decisions, since these require a more nuanced and up-to-date picture than national-level indicators can provide. However, serving this demand appropriately remains a challenge in developing countries: censuses are collected every ten years at best for a limited scope of indicators. Household surveys are usually conducted more frequently covering a broad range of indicators, although they may not provide representative results at higher levels of disaggregation. In addition, data collection costs for both instruments scale unfavourably with the sample size. While some countries can link statistical data with high-quality administrative records in order to fill gaps, e.g. using small area estimation methods (\cite{Pfe13,RaoMol15,Tzavidis2018FromStatistics}), many others lack appropriate administrative data systems to do so.

To overcome the shortage of geographically disaggregated data, academic literature has seen a rise of approaches in recent years that use non-traditional (big) data sources. 
In particular, night-time lights and land use information captured by satellites, as well as mobility and social network insights via mobile phone metadata, are increasingly used for estimating disaggregated key statistical indicators related to economic growth (\cite{Chen2011UsingStatistics, Henderson2012MeasuringSpaceb,  Pinkovskiy2016LightsDebate}), literacy (\cite{ Sundsy2016CanCountry, Schmid2017ConstructingSenegalb}), population density (\cite{Harvey2002EstimatingLimitations,  Bonafilia2019MappingLearning, Leyk2019AllocatingUseb, Steinnocher2019EstimatingImagery}) or poverty (\cite{Jean2016CombiningPoverty, Pokhriyal2017, Weidmann2017UsingWealth}). The United Nations (UN) recommends using satellite imagery to prioritize and check geospatial processes such as the delineation of enumeration areas during census preparation (\cite{DepartmentofEconomicandSocialAffairsUN2009HandbookActivitiesc}). It further supports the construction of population grids as a common spatial reference system as proposed by \cite{Stevens2015DisaggregatingData, Freire2016DevelopmentResolution, Boo2020ASurveys}.

While these approaches can increase spatial granularity by assuming that correlations captured for large areas also hold for small areas, they suffer of five major shortcomings: First, they still require disaggregated statistical data from surveys or censuses for training and model fitting, ideally captured at the same point in time. Second, only variables of interest that correlate with covariates from the novel data source can be modelled appropriately. Third, little research has been done so far to investigate whether these inferential relationships also hold over time. Fourth, some of the novel data sources are privately owned, e.g. high resolution satellite imagery and mobile phone metadata. Creating reliable long-term access options, a precondition for embedding them in statistical business processes, is an institutional and legal challenge on its own. Fifth, many new approaches use sophisticated machine learning in combination with data generated in processes beyond the control of statistical offices. Adopting them would require a major paradigm shift in the work of official statisticians. However, and most likely given the capacity constraints many statistical offices in the world face, it discourages adoption altogether.

 Another strain of research, usually known as updating, intercensal or postcensal estimation methods, focuses on the periodicity of statistical data by tackling the problem of the time-gap between two sources of information. Traditional demographic techniques such the component method (\cite{united1956manual}) and the vital rates method or regression symptomatic procedures (\cite{Rao03}) build on long-term trends around fertility, mortality, and migration and have been extensively used to obtain postcensal population estimates, often called population projections. Most of these techniques rely on survey information, therefore limiting the level of spatial granularity for which population updates can be provided. Large-area population estimates are then distributed to small areas via the proportion method (\cite{united1956manual}), i.e. fixed population shares captured in the latest census. Population shifts within these large areas, e.g. induced by urbanization, desertification or internal conflicts are thus not accounted for. Hence, the lack of disaggregated, frequent, relevant, reliable, and timely statistics is still largely unsolved.

The structure preserving estimation (SPREE) method has been widely used to produce intercensal estimates for disaggregated population counts (\cite{Purcell1980PostcensalDomains}). Unemployment (\cite{luna2015small}), occupational classification (\cite{Hid09}) and poverty estimation (\cite{Isidro2016ExtendedPovertyb}) are examples of the application of this method and its extensions. In SPREE, disaggregated postcensal estimates are generated by updating the margins (called the \textit{allocation structure}) of a census composition (called the \textit{association structure}). The updated margins, i.e. the distribution of the variable of interest (e.g. poverty status by sex) and the population sizes for each domain, are usually provided by recent survey data. In case population sizes by domains cannot be sufficiently generated due to sampling design constraints, population projections as described above can be used instead (\cite{luna2016multivariate}).

In this paper, we provide a unified framework for the two aforementioned, so far mainly disparate, fields of research - using novel data sources to add spatial granularity and updating existing census structures with long-term trends and recent data to add disaggregation, periodicity, and reliability. Specifically, we study the potential benefit of including auxiliary data in the SPREE process by strengthening the population estimates in local areas. The aim is relaxing rigid assumptions on the subnational population allocation between census rounds by tracking population changes across small areas over time using auxiliary information without jeopardizing the compelling simplicity of the SPREE approach. We also provide uncertainty estimates for the postcensal predictions based on our approach following a semiparametric bootstrap procedure. We apply the proposed methodology in a case study in Senegal. Since its latest census in 2013, Senegal has started to roll out a nation-wide social safety net program to lift families out of poverty including cash transfers for up to every fourth Senegalese household (\cite{GouvernementduSenegal2017ProgrammePNBSF,ANSD2015EnqueteVulnerables1}). Therefore, we expect that the only source of commune-level data on multidimensional poverty, the census 2013 data, has become quickly outdated and therefore is ill-suited for further local policy guidance. In a first step, we validate the value added of our proposed methodology by updating the census 2002 to 2013 and comparing the results with the actual census 2013. Finally, using data from census 2013, multiple rounds of major household surveys, population projections, and satellite imagery-based population estimates we retrieve annual point predictions including corresponding mean squared error (MSE) estimates of the multidimensional headcount ratio for the 552 communes in Senegal up to 2020.

The document has the following structure: a summary of SPREE is included in Section 2 as well as the proposed updating strategy which incorporates auxiliary information such as satellite imagery. MSE estimates for corresponding point estimates are also provided. Section \ref{section-case} is dedicated to introduce the case study and the data sources used in both the validation and the application. In Section \ref{section-validation} we present the simulation exercise to validate our proposal. Section \ref{section-application} summarises the application results in which we provide small-area-level updates of the multidimensional headcount ratio for several postcensal years in Senegal. Conclusions and further research questions are presented in Section \ref{section-conclusion}.

\section{Methodology}\label{section-methodology}

The method proposed to obtain updated indicators in small areas by including auxiliary information is described in this section. After a brief introduction to SPREE, we present the main scientific contribution of this paper: incorporating auxiliary information to improve the accuracy of small-area census updates by relaxing the assumption of fixed population margins between census years.

\subsection{Structure Preserving Estimation (SPREE)}\label{subsection-spree}

SPREE is an updating technique for intercensal population compositions. In general, population compositions can be thought of as multiway contingency tables. We denote the composition of interest as $Y_{aj,t}$ with $a=1,\dots,A$ representing the target areas i.e. small areas and $j =1,\dots,J$ representing the categories of interest, e.g. poor/non-poor, at time $t=1,\dots,T$. Consequently, $Y_{a,t}$ represents the row margins, i.e. population sizes for each area, and $Y_{j,t}$ the column margins, i.e. distribution of the categories of interest at national level (global totals) since the last census. Here, we assume the initial population composition to be generated at time $t=0$. Since that is usually the year of the latest census, we simply call it the \textit{census composition}. For better legibility, we denote the census year without a time index, e.g. $Y_{aj}$ instead of $Y_{aj,0}$. Further, we assume that smaller areas are disjoint subsets of larger areas (as it is commonly the case for administrative areas) which we denote with $k=1,\dots,K$.

In order to illustrate how a census composition can be updated via the margins, we describe it as a log-linear model (here for a two-way case such as poverty status by area):
\begin{equation*}
    \log{Y_{aj}}= \alpha^Y_0 + \alpha^Y_a + \alpha^Y_j + \alpha^Y_{aj} ,
\end{equation*}
where $\alpha^Y_0$ represents the overall mean, $\alpha^Y_a$ the main effect of areas, $\alpha^Y_j$ the main effect of categories, and $\alpha^Y_{aj}$ the interaction term also known as the association structure. The definition of these parameters can be found in detail in most of the SPREE literature (see e.g. \cite{luna2016multivariate}). Following the original proposal of \cite{Purcell1980PostcensalDomains}, two basic structures are required to apply this methodology: an association structure $\alpha^Y_{aj}$ which provides information about the interaction between areas (rows) and categories (columns), and an allocation structure represented by $\alpha^Y_0$, $\alpha^Y_a$ and $\alpha^Y_j$ that provides information about area and category terms. Generally, in order to capture the association structure appropriately, each combination of areas and categories (called \textit{cells}) must be backed by sufficient data. Especially for countries lacking effective administrative data systems, censuses are the only source of information at that level of detail as surveys usually suffer of sample size constraints. Consequently, the census composition is only available every ten years at best for many countries in the world and therefore the association structure is assumed to be constant between census years $\big(\alpha^Y_{aj,t} = \alpha^Y_{aj} \big)$. Usually the allocation structure is gathered from direct survey estimates, however, spatial disaggregation here is often limited by the lack of reliable subnational population margins.

The fitting process in the original SPREE model is achieved via the iterative proportional fitting (IPF) algorithm (\cite{Deming1940}), where reliable survey row $\big(\hat{Y}_{a,t}\big)$ and column margins $\big(\hat{Y}_{j,t}\big)$ are fitted to a census composition ($Y_{aj}$). Then, the iterative process to obtain the updated point estimates is indicated as in \cite{luna2016multivariate} as:
\begin{equation}\label{eq:point}
  \hat{Y}_{aj,t} = \text{IPF} \big(Y_{aj},\hat{Y}_{a,t}, \hat{Y}_{j,t}\big).
 \end{equation}
 
There are some basic limitations to this methodology: 
\begin{enumerate}[label=(\alph*)]
    \item the survey(s) used to generate $\hat{Y}_{j,t}$ and $\hat{Y}_{a,t}$ must contain the same variable definition as the census or a high correlate of it (\cite{Green1998}),
    \item the survey margins must follow the same disaggregation scheme as defined in the census (areas/categories),
    \item the margin totals $\sum_{a=1}^{A} \hat{Y}_{a,t}$ and $\sum_{j=1}^{J} \hat{Y}_{j,t}$ need to add up perfectly, 
   \item only categorical variables are allowed (e.g. poverty status = poor/not poor).
\end{enumerate}

The literature on SPREE methodology can be roughly summarised into three groups: a) original SPREE as defined by \cite{Purcell1980PostcensalDomains} and the proposal of \cite{Noble2002} to formulate SPREE under the generalized linear model (GLM) framework, b) SPREE extensions focused on bias reduction as suggested by \cite{Zhang2004SmallCross-classifications} and later extended by \cite{luna2016multivariate} through incorporating cell-specific random effects and informative sampling design  and c) SPREE extensions to account for situations where the variable of interest is not available in the census (\cite{Isidro2016ExtendedPovertyb}). As far as we know, the only work that uses additional sources of information besides one census, survey data, and population projections is \cite{luna2015small} who consider a second census composition for updating ethnicity categories by age groups in England in 2011. While the population census provided information for all age groups, a scholar census offered better information for children between five and 15 years old. The authors studied the possibilities of updating the estimates considering that one of the age groups has two different data sources available. Nonetheless, the authors did not observe improvements by using this additional source of information. However, they point to the need for further research to better understand the potential value added of auxiliary information in SPREE. Moreover, to the best of our knowledge, all studies in this field assume the presence of reliable and up-to-date population margins for the areas of interest, which may not be ubiquitously available in many small-area settings. 

\subsection{SPREE with auxiliary information}

Our contribution here is straight-forward: as demographic projections and surveys
usually provide reliable population totals only for large areas ($\hat{Y}_{k,t}$), a common-place method uses fixed population proportions $p_{k_a}$ captured in $t=0$ for distributing large-area counts to small areas:
\begin{equation}\label{eq-prop-pbar}
\hat{Y}_{a,t} =  \hat{Y}_{k,t} \cdot p_{a} \quad \text{with} \quad \sum_{a=1}^{A} p_{a} = \sum_{a=1}^{A}\frac{Y_{a}}{Y_{k}} = 1  \quad \forall \quad a \subseteq k.
\end{equation}

We relax this obviously strong assumption that subnational population shares remain fixed between censuses by using auxiliary information to track population changes dynamically for small areas over time. Consequently, we replace $p_{a}$ with $\hat{p}_{a,t}$, so Equation \ref{eq-prop-pbar} becomes: 
\begin{equation}\label{eq-prop-phat}
\hat{Y}_{a,t} = \hat{Y}_{k,t} \cdot \hat{p}_{a,t} \quad \text{with} \quad \sum_{a=1}^{A}\hat{p}_{a,t} = \sum_{a=1}^{A}\frac{\hat{Y}_{a,t}^{\text{Aux}}}{\hat{Y}_{k,t}^{\text{Aux}}} = 1  \quad \forall \quad a \subseteq k,
\end{equation}

where $\hat{Y}_{a,t}^{\text{Aux}}$ and $\hat{Y}_{k,t}^{\text{Aux}}$ represent the population sizes of small areas $a$ and large areas $k$ at time $t$ estimated using auxiliary data. 
Notice that it may be possible to rely on $\hat{Y}_{a,t}^{\text{Aux}}$ only to estimate $\hat{Y}_{a,t}$. However, demographic projections and surveys capture information on population whereabouts for large areas in a statistically rigorous and reliable manner and are therefore used as regional-level benchmarks through dasymetric mapping (\cite{Stevens2015DisaggregatingData}). This allows us to combine the benefits from both auxiliary data and statistical data. First, reliable counts are distributed to smaller areas not covered so far. Second, only information on the within-area distribution of counts are used from auxiliary information, not on the totals. This makes SPREE for small areas less prone to quality issues in the auxiliary data.

Besides administrative records, big data sources such as mobile phone metadata and satellite imagery appear to be promising auxiliary information in SPREE. For satellite imagery, five characteristics make it a suitable candidate for that: Satellite imagery provides virtually 1) global coverage at 2) high spatial resolutions for 3) frequent time intervals on 4) human-made impact in a 5) structured and - at least to some extent - harmonised way. In regard to population changes, built settlement growth, and changes in night-time lights may constitute the potentially most relevant indicators on human impact from satellite imagery. Data products from remote sensing are currently rapidly evolving in terms of quality and user-friendliness. Examples for open-access data products from remote sensing are the
Copernicus program (\cite{Union2020Copernicus:Earth}) and \cite{WorldPopwww.worldpop.org-SchoolofGeographyandEnvironmentalScienceUniversityofSouthamptonDepartmentofGeographyandGeosciencesUniversityofLouisvilleDepartementdeGeographie2018WorldPopb1}.

\subsection{Assessment of uncertainty}\label{subsection-uncertainty}

The original SPREE approach assumes that the association structure is fixed and therefore only the allocation structure provides a variation component. While \cite{Purcell1980PostcensalDomains} propose linearization or replication techniques to account for the introduced uncertainty, \cite{rao1986synthetic}, \cite{isidro2010intercensal} and \cite{luna2016multivariate} provide also analytical approximations for the variance of the estimator and other measures of uncertainty for SPREE and its extensions. In contrast, the data generating process of auxiliary data in general and specifically the population margins derived from satellite imagery is usually not known and it has to be expected that uncertainty in these data sources follows complex, non-trivial patterns. For example, satellite imagery run through multiple stages of data processing with dependencies to other data sources such as OpenStreetMaps (\cite{Stevens2015DisaggregatingData}) before providing population estimates. To some extent, this complexity also holds true for uncertainty in demographic projections.
As we assume the uncertainty stemming from external sources like satellite imagery to be non-negligible, but also non-trivial, we opt to follow the replication idea proposed by \cite{Purcell1980PostcensalDomains} which is also the basis of some MSE alternatives suggested in \cite{luna2016multivariate} instead of pursuing an analytical strategy that may require strong assumptions on some of the sources of uncertainty. Therefore, we propose an empirical approximation of the uncertainty based on a semiparametric bootstrap in order to give some indication on the trustworthiness of a point estimate while acknowledging that it most likely captures only a fraction of the total variation.
Since the allocation structure is constructed based on auxiliary data, population projections and survey data, a fully parametric approach is not straightforward. For this reason and to avoid misspecification issues, the column and row margin are resampled in an non-parametric way, where the column margin is provided by direct survey estimates and the row margin is derived from auxiliary information. Following existing literature (see e.g. \cite{luna2016multivariate}), census data is replicated following a multinomial distribution. In addition, we account for the stochastic nature of the overall population by resampling the population of each area using a Poisson distribution. This step is complementary and intended for use cases where census counts are subject to uncertainty, e.g. due to quality issues or the availability of only a census sample. To summarize, the steps to obtain the MSE of the updated estimates $\hat{Y}_{aj,t}$ via a mixed semiparametric bootstrap are as follows:

\begin{enumerate}[label=(\alph*)]
   \item obtain the point estimate $\hat{Y}_{aj,t}$ from Equation \ref{eq:point} using the row margin $\hat{Y}_{a,t}$ as defined in Equation \ref{eq-prop-phat} and compute  $\hat\pi_{aj,t} =  \frac{\hat{Y}_{aj,t}}{\hat{Y}_{a,t}}$,
   \item for $b = 1, \dots, B$:
   \begin{enumerate}[label=(\roman*)]
    \item create population composition $\hat{Y}_{aj,t}^{\text{Mult},b}$ following a      multinomial distribution, 
    denoted by $\text{Multinom}\Big(\hat{Y}_{a,t}^{\text{Pois},b},\hat\pi_{aj,t}\Big)$, where the row margins $\hat{Y}_{a,t}^{\text{Pois},b}$ are obtained assuming a Poisson distribution, defined by $\text{Pois}\big(\lambda = \hat{Y}_{a,t}\big)$, 
  \item generate auxiliary-based row margin $\hat{Y}_{a,t}^b$ by resampling from $\hat{Y}_{a,t}$,
  \item generate survey-based column margin $\hat{Y}_{j,t}^b$ by resampling from $\hat{Y}_{j,t}$,
  \item apply SPREE (Equation \ref{eq:point}) for each replicate, generating $B$ updated compositions $\hat{Y}_{aj,t}^b = \text{IPF}\Big(\hat{Y}_{aj,t}^{\text{Mult},b}, \hat{Y}_{a,t}^b, \hat{Y}_{j,t}^b\Big)$,  
\end{enumerate}
    \item the MSE of $\hat{Y}_{aj,t}$ is estimated as: 
    \begin{equation*}
 \widehat{MSE}\big(\hat{Y}_{aj,t}\big) =  \frac{1}{B}  \sum_{b=1}^{B} \Big(\hat{Y}_{aj,t}^b - \hat{Y}_{aj,t}^{\text{Mult},b}\Big)^2.
    \end{equation*}
  \end{enumerate}

\section{Case study: multidimensional poverty in Senegal}\label{section-case}

In 2013, the Senegalese government launched a nation-wide family security benefits program as part of a wider social safety net initiative. Officially termed \textit{Programme National de Bourses de Sécurité Familiale} (PNBSF), the goal of the ongoing program is `to contribute to the fight against the vulnerability and social exclusion of families through integrated social protection in order to promote their access to social transfers and to strengthen, among other things, their educational, productive and technical capacities' (\cite{GouvernementduSenegal2017ProgrammePNBSF}).
Its main component are cash transfers to disadvantaged families that in turn commit to schooling, keeping vaccination records and registering newborns and children in civil status (\cite{ANSD2015EnqueteVulnerables1}). The cash transfers (100,000 FCFA, i.e. approximately 180 USD per year and household) are accompanied by coordination, communication, and evaluation activities. After a one-year pilot phase with 50,000 poor or vulnerable Senegalese households in 2013, the program was scaled up to 250,000 households in 2014 and onwards. A further extension to a total of 400,000 households - that is approximately every fourth household in Senegal - is currently underway. Consequently, it can be expected that poverty-related measures for small areas in Senegal will have become quickly outdated since the latest census in 2013. Even though the program efficiency is monitored including a large-scale baseline survey considering treatment and control group, this does not provide a representative view on socio-economic development in Senegal in general and for sub-groups in the 552 communes of Senegal, specifically. 

The aim of the case study is to provide annual census updates of multidimensional poverty for small areas in Senegal until 2020 to inform about main drivers of multidimensional poverty for a specific sub-group: individuals living in female-headed households. To make sure our methodology adds value vis-à-vis existing approaches, we validate our proposal using censuses from 2002 and 2013. As we combine multiple sources of data in this case study, we first introduce them together with the target indicator - the headcount ratio of the multidimensional poverty index - in this section before presenting the validation results in Section \ref{section-validation} and the application results in Section \ref{section-application}. 
The full code for this paper is available on GitHub:  \url{https://github.com/tilluz/Auxiliary-SPREE}

\subsection{Geography of Senegal}\label{section-geography}

\begin{figure}[ht!]
    \centering
    \includegraphics[width=\textwidth]{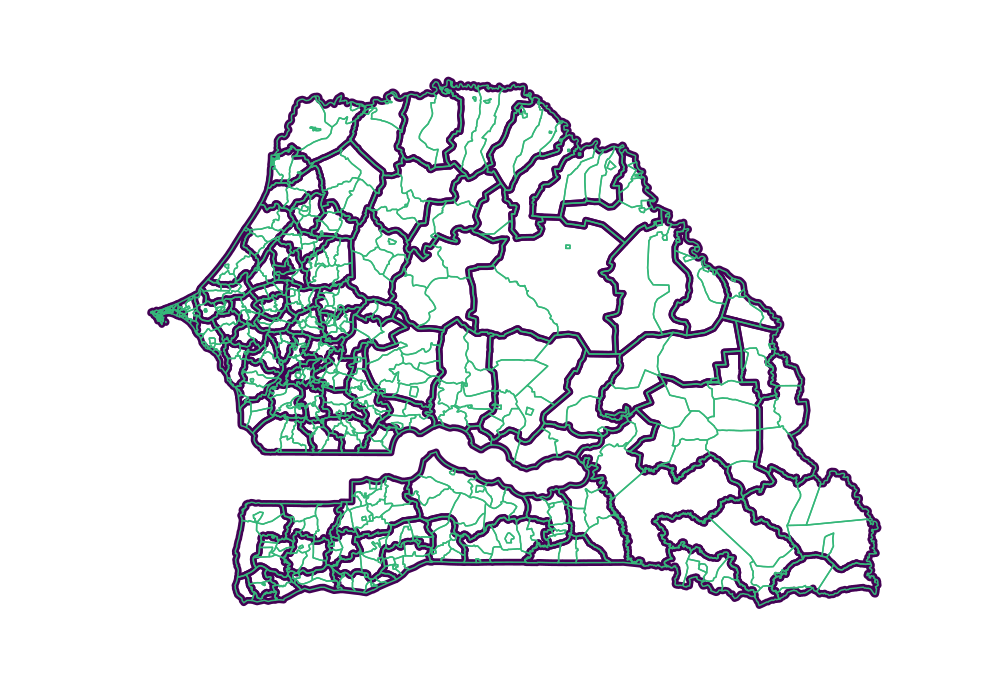}
    \caption{\textbf{Overlay of RGPH 2002 (grey) and the RGPHAE 2013 (green).}\\
    The 552 communes from the RGPHAE 2013 are perfect subsets of the 103 arrondissements of the RGPH 2002.}
    \label{fig:sim-overlay}
\end{figure}

Senegal is administratively divided in 14 \textit{régions} (NUTS1), 45 \textit{départements} (NUTS2), 103 \textit{arrondissements} (NUTS3) and 552 \textit{communes} (NUTS4). The total population was at 13.5 million in 2013 and projected to be at 16.2 million in 2020. We choose the 552 communes as target domains in the application as they represent the level of administration responsible for the provision of public services in the decentralization efforts set out in the national development plan of Senegal (Plan Sénégal Émergent) (\cite{GouvernementduSenegal2014PlanEmergent}). Since a number of land reforms have taken place in Senegal between 2002 and 2013, the smallest common denominator in terms of administrative boundaries for analysis between 2002, and 2013 is the NUTS3 level (103 arrondissement) of  the census 2002 \textit{Recensement Général de la Population et de l’Habitat} (RGPH 2002). For evaluation, we consequently aggregate the data of the census 2013 \textit{Recensement Général de la Population et de l’Habitat, de l’Agriculture et de l’Elevage} (RGPHAE 2013) from the 552 communes to the 103 arrondissements. Figure \ref{fig:sim-overlay} shows an overlay of the two administrative structures.

\subsection{Multidimensional poverty in Senegal}\label{subsection-mpi-senegal}

The global multidimensional poverty index (MPI) for Senegal based on \cite{Alkire2019The2019} is  calculated considering the incidence of poverty $H$ and the average deprivation share $D$:  
$$\text{MPI}= H  \times  D.$$
The MPI varies from 0 to 1 on a continuous scale, where 1 represents the highest level of MPI poverty. As SPREE requires the target indicator to be categorical, we focus on the incidence $H$ (also called \textit{MPI headcount ratio}) in this case study. The MPI headcount ratio is defined as the proportion of people identified as multidimensional poor. It is calculated based on the (weighted) number of deprivations that they experience (also called \textit{deprivation score}). The MPI is composed of ten indicators measuring different dimensions of deprivation. A household with a deprivation score of at least 1/3 is considered multidimensional poor. The deprivation indicators $I_i$ with the respective weight $w_i$ are summarized in Table \ref{tab:mpi-overview}.

\begin{table}[ht]
\centering
\small{
\begin{tabular}{llcc}
\hline
Dimension & Indicator ($i$) & Weight ($w_i$) &Available\\
\hline
Health 	& Nutrition & 1/6 &No\\
&Child mortality* & 1/6  &Yes \\
\hline
Education & Years of schooling &1/6 &Yes\\
& School attendance &1/6  &Yes\\
\hline
Living standards & Cooking fuel &1/18 &Yes \\
& Sanitation &1/18  &Yes\\
&Drinking water & 1/18  &Yes\\
& Electricity & 1/18  &Yes\\
& Housing & 1/18 &Yes\\
& Assets &1/18  &Yes\\
\hline
\end{tabular}
\caption{\textbf{Dimensions and weights for measuring the MPI.}\\
*As data on `nutrition' is not available, `child mortality' receives the full weight of the `health' dimension (1/3).}
\label{tab:mpi-overview}
}
\end{table}

The percentage contribution $\phi$ of indicator $i$ to the overall MPI headcount ratio $H$ is defined as:
\begin{equation*}
    \phi_i = w_i*\frac{H_i}{H}.
\end{equation*}

\begin{figure}[ht!]
\captionsetup[subfigure]{justification=centering}
    \centering
    \includegraphics[width=\textwidth]{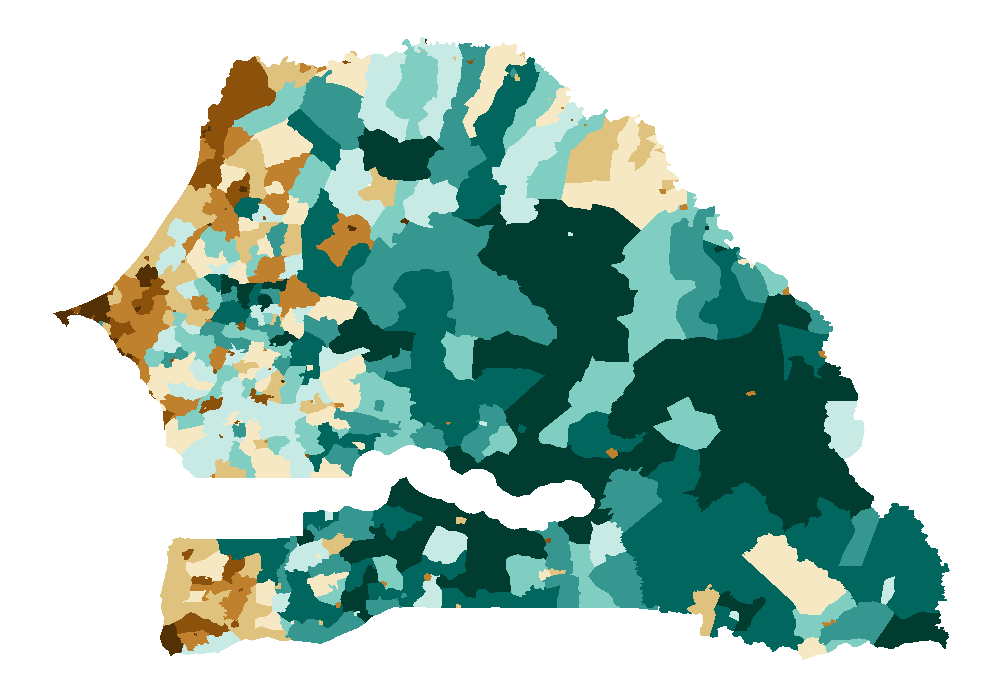}
    \caption{\textbf{MPI headcount ratio in 2013 by commune.}\\
    The unweighted commune-level MPI headcount ratio in Senegal in 2013 averages at 0.75 (white) with a minimum at 0.1 (dark brown) and a maximum at 1 (dark green).}
    \label{fig:sen-hcr}
\end{figure}

Indicator-specific contributions to the overall MPI headcount ratio are particularly interesting for policy-making as they rank drivers of multidimensional poverty along the usual responsibilities of line ministries (e.g. education, energy, health, infrastructure etc.) in a simple-to-visualize and easy-to-understand manner. For more details, we refer to \cite{Alkire2019The2019}. Although the survey data used in this application provide information for all MPI indicators, the census data lacks of the `nutrition' indicator from the `health' dimension. Even though the situation when there is lack of information in the census can be addressed within SPREE as shown by \cite{Isidro2016ExtendedPovertyb}, we focus on the nine remaining indicators for the sake of simplicity. Therefore, the indicator `child mortality' in the `health' dimension receives a weight of 1/3 instead of 1/6 (see Table \ref{tab:mpi-overview}).
The multidimensional headcount ratio ($H$) in 2013 is shown for subgroups by région in Table \ref{tab:census-h} and by commune in Figure \ref{fig:sen-hcr}.

\begin{table}[ht]
\centering
\small{
\begin{tabular}{lcc|lcc}
		\hline
		Région & $H$ & $H_{\text{female}}$ & Région & $H$ & $H_{\text{female}}$\\ 
		\hline
Dakar & 51.8 & 50.3 & Louga & 69.9 & 57.4 \\ 
  Diourbel & 69.7 & 58.5 & Matam & 80.6 & 70.7 \\ 
  Fatick & 75.8 & 62.9 & Saint-Louis & 68.6 & 59.7 \\ 
  Kaffrine & 81.9 & 65.5 & Sedhiou & 84.6 & 74.2 \\ 
  Kaolack & 72.6 & 58.4 & Tambacounda & 85.6 & 71.8 \\ 
  Kedougou & 81.8 & 72.7 & Thies & 64.2 & 55.8 \\ 
  Kolda & 82.7 & 65.9 & Ziguinchor & 65.3 & 61.1 \\ 
		\hline
	\end{tabular}
	\caption{\textbf{Regional-level MPI headcount ratios by sub-group from 2013 census}.\\
	Headcount ratios are given for the overall population ($H$) and the population living in a female-headed household ($H_{\text{female}}$).}
	\label{tab:census-h}
	}
\end{table}

\subsection{Data sources}\label{subsection-data}

\subsubsection{Recensement Général de la Population et de l'Habitat, de l'Agriculture et de l'Elevage (RGPHAE) 2013}\label{subsubsection-appl-rgphae}

Data collection for the latest census took place between November 19th, 2013 and December 14th, 2013. Unit-level data of the population and housing census with 10\% of the households sampled within each of the 17,145 census districts is provided for scientific use via the microdata portal of the \textit{Agence Nationale de la Statistique and de la Démographie} (ANSD) - the Senegalese national statistical office. The RGPHAE 2013 provides data for all MPI indicators but not for `nutrition'. Some indicators suffer of item nonreponse that apparently do not follow a missing completely at random (MCAR) scheme. Table \ref{tab:rgphae-overview} provides an overview over the missing data treatment in the RGPHAE 2013, where can be seen that the population part of RGPHAE 2013 exhibits large-scale item-nonresponse. The strong imputation effects in these indicators are somehow expected due their respective definition. A household is deprived in the indicator `child mortality' if any child has died in the household. Thus, the higher the coverage of individuals within the household, the more likely it is to fulfill the criteria of deprivation. Same holds for the indicator `school attendance', which defines a household as deprived if any school-aged child is not attending school up to class eight. The contrary holds for the indicator `years of schooling', which defines a household as deprived if no household member has completed six years of schooling. The more household members are covered, the more likely it is for a household not to fulfill the criteria of deprivation.
We address the item nonresponse by imputing missing values in a single imputation using chained equations (see e.g. \cite{White2011MultiplePractice}). Even though we acknowledge that imputation introduces additional uncertainty, we do not explicitly account for it as this goes beyond the scope of the paper. Consequently, we put it as further research.

\begin{table}[ht]
\centering
\small{
		\begin{tabular}{lccc}
		\hline
		Indicator & \% missing  & \% deprived & \% deprived (imputed)\\
		\hline
		Nutrition & - & - & - \\
		Child mortality & 65.4  & 27.9 & 53.6\\
		\hline
 		Years of schooling & 60.6 & 15.1 & 11.3 \\
 		School attendance & 60.6 & 40.8 & 57.4 \\
 		\hline
		Cooking fuel & 0.0 & 65.8 & 65.8 \\
		Sanitation & 0.0 & 30.8 & 30.8 \\
		Drinking water & 0.0 & 15.7 & 15.7 \\
		Electricity & 0.0 & 38.9 & 38.9 \\
		Housing & 0.0 & 26.2 & 26.2 \\
		Assets & 0.0 & 16.6 & 16.6 \\
		\hline
	\end{tabular}
	\caption{\textbf{Missing MPI data and imputation effects in the RGPHAE 2013.}\\}
	\label{tab:rgphae-overview}
	}
\end{table}

\subsubsection{Recensement Général de la Population et de l'Habitat (RGPH) 2002}\label{subsubsection-appl-rgph}

Similar to RGPHAE 2013, the \textit{Recensement Général de la Population et de l'Habitat 2002} (RGPH 2002) data is available at the unit-level with 10\% of the households sampled within each of the census districts for scientific use via the microdata portal of ANSD. As with RGPHAE 2013 data, missing values are treated by single imputation using chained equations.

\subsubsection{Household survey data}\label{subsubsection-appl-dhs}

In this case study, we use data from Senegalese Demographic and Health Surveys (DHS) collected for the years 2013 and beyond. The DHS program is mainly funded by the United States Agency for International Development (USAID) with the aim of provide accurate and reliable indicators of population, health, and nutrition at national, residence (urban-rural) and regional levels (NUTS1). Details on the respective survey rounds are summarized in Table \ref{tab:dhs-overview}.  

\begin{table}[ht]
\centering
\small{
		\begin{tabular}{ccccc}
		\hline
		Year & Contract phase & Data collection & Sample size & Clusters  \\
		\hline
		2013	 	& VI & Jan-Oct & 4 174 & 200 \\
		2014	 	& VII & Jan-Oct & 8 406 & 400 \\
		2015	 	& VII & Jan-Oct & 4 511 & 214\\
		2016	 	& VII & Jan-Oct & 4 437 & 214\\
		2017		& VII & Apr-Dec & 8 380 & 400\\
        2018		& VIII & Jun-Dec & 4 592 & 214\\
        2019		& VIII & Apr-Dec & 4 538 & 214\\
			\hline
	\end{tabular}
	\caption{\textbf{Description of the DHS data set available.}}
	\label{tab:dhs-overview}
	}
\end{table}

\begin{figure}[ht!]
\captionsetup[subfigure]{justification=centering}
    \centering
    \includegraphics[width=\textwidth]{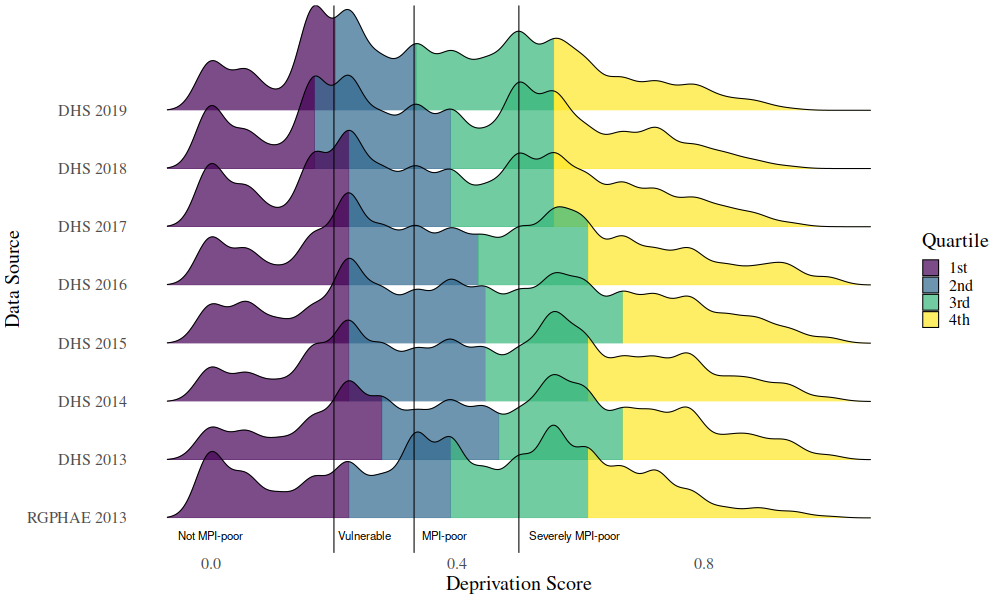}
    \caption{\textbf{Distribution of household-level deprivation scores by data source.}\\
    Cutoffs are 0.2 (vulnerable to multidimensional poverty), 0.33 (MPI-poor) and 0.5 (severely MPI-poor), respectively.}
    \label{fig:score_distribution}
\end{figure}

Indicator definitions and/or response options slightly differ between DHS surveys and RGPHAE 2013 leading to differences in the results as shown in Figure \ref{fig:score_distribution} and Table \ref{tab:ind-overview}. For example, while the DHS surveys provide data on the time travelled for safe drinking water, RGPHAE 2013 does not. Therefore, we do not account for the time travelled in the respective indicator even though it has a major impact on the deprivation score for drinking water in Senegal. Nevertheless, regional-level correlations between indicators of the DHS 2013 and RGPHAE 2013 are generally quite strong with an average of 0.68 ranging from 0.40 for `assets' to 0.99 for `cooking fuel'.

\begin{table}[ht]
\begin{center}
	\centering
\small{
	  \begin{tabular}{l|cccccccc}
        \toprule
    \multirow{2}{*}{Indicator} &\multicolumn{1}{c}{RGPHAE} & \multicolumn{7}{c}{DHS}\\ 
        \cmidrule{2-9} 
        & 2013 & '13 & '14 & '15 & '16 & '17 & '18 & '19 \\ 
         \midrule
Child mortality & 53.9  & 41.4  & 39.9  & 39.4  & 36.0  & 37.6  & 36.1  & 32.2  \\ 
  Years of schooling & 10.0  & 36.2  & 36.1  & 40.0  & 38.6  & 33.0  & 34.8  & 37.0  \\ 
  School attendance & 64.4  & 80.8  & 79.7  & 74.4  & 76.5  & 76.8  & 76.8  & 77.9  \\ 
  Cooking fuel & 75.8  & 81.0  & 78.7  & 79.0  & 74.0  & 22.5  & 20.9  & 21.3  \\ 
  Sanitation & 35.2  & 42.1  & 41.1  & 45.4  & 41.4  & 36.5  & 30.5  & 31.8  \\ 
  Drinking water & 17.5  & 24.1  & 21.7  & 25.2  & 23.2  & 22.5  & 21.9  & 23.4  \\ 
  Electricity & 43.3  & 47.0  & 43.3  & 41.7  & 37.0  & 38.3  & 33.8  & 29.4  \\ 
  Housing & 27.9  & 26.5  & 26.9  & 22.4  & 22.5  & 21.8  & 14.8  & 15.1  \\ 
  Assets & 12.4  & 11.5  & 11.0  & 12.5  & 12.4  & 11.5  & 10.0  & 10.8  \\ 
  MPI headcount & 68.4  & 69.0  & 67.9  & 67.5  & 64.7  & 59.3  & 58.1  & 57.8  \\ 
        \bottomrule
    \end{tabular} 
    \caption{\textbf{Overview over indicator-specific headcount ratios by data source.}\\}
	\label{tab:ind-overview}
	}
	\end{center}
\end{table}

Differences between the categories available to create the MPI indicators lead also to 
differences in the distribution of the deprivation scores when comparing RGPHAE 2013 and the DHS 2013. For instance, while the census asks whether public sanitation facilities are used, the DHS asks whether sanitation facilities are shared (which could also occur privately among a small number of households).

\subsubsection{Demographic projections}\label{subsubsection-appl-proj}

Demographic projections inform about the possible future demographic composition of a country. Usually, demographic projections take the latest census as base population and use hypotheses on the future development of fertility, mortality and migration in a country. In Senegal, demographic projections are produced by ANSD disaggregated by 5-year age groups, gender, region, and urban status based on RGPHAE 2013 using the software SPECTRUM following the cohort component method (\cite{united1956manual}): 
\begin{equation*}
  \text{pop}_{t1} = \text{pop}_{t_0} + (\text{births}_{t_1} - \text{deaths}_{t_1}) + (\text{immigration}_{t_1} - \text{emigration}_{t_1}).
 \end{equation*}
Population projections are made on the regional level, which are then distributed via fixed population shares determined during the census year to lower administrative levels (proportion method (\cite{united1956manual})).
As reliable regional-level data on migration is not available, the migration pattern captured in the census is treated as static over time. For estimating current fertility and death trends, the ANSD uses regional-level survey data, usually from the latest DHS. For details on the methodology applied to create population projections in Senegal, we refer to \cite{AgenceNationaledelaStatistiqueetdelaDemographie2013Rapport2013-20631}.

\subsubsection{Satellite Imagery}\label{subsubsection-appl-si}

\begin{figure}[ht!]
\captionsetup[subfigure]{justification=centering}
     \centering
     \begin{subfigure}[b]{0.49\linewidth}
         \centering
         \includegraphics[width=\textwidth]{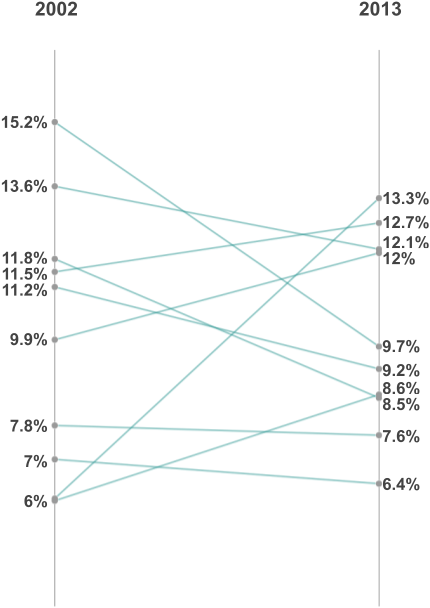}
         \caption{RGPHAE 2013}
         \label{fig:sim_slope_true}
     \end{subfigure}
     \begin{subfigure}[b]{0.49\linewidth}
         \centering
         \includegraphics[width=\textwidth]{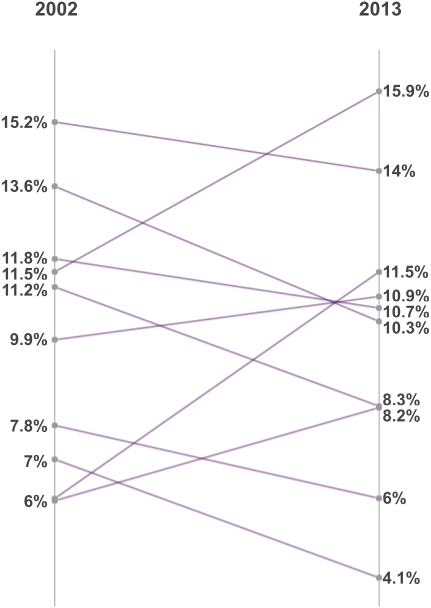}
         \caption{Satellite imagery-based population estimates}
         \label{fig:sim_slope_ppp}
     \end{subfigure}
        \caption{\textbf{Development of the small-area population shares between 2002 and 2013 for the Dakar region.}\\
        The within-region population shares always add up to 100\%.}
        \label{fig:sim_slope}
\end{figure}

In this case study, we use satellite imagery provided by \cite{WorldPopwww.worldpop.org-SchoolofGeographyandEnvironmentalScienceUniversityofSouthamptonDepartmentofGeographyandGeosciencesUniversityofLouisvilleDepartementdeGeographie2018WorldPopb1} for estimating dynamic subnational population shares on a frequent basis. The idea behind this is to account for population shifts occurring within regions that are so far not captured via the proportion method. By using the Dakar région as example, Figure \ref{fig:sim_slope} shows the evolution of the population shares from 2002 to 2013. Figure \ref{fig:sim_slope_true} shows how the population shares have changed based on population counts from RGPH 2002 and RGPHAE 2013, and Figure \ref{fig:sim_slope_ppp} shows changes in the population shares based on RGPH 2002 and satellite imagery-derived population estimates from WorldPop. The proportion method (not depicted) assumes fixed population shares, i.e. horizontal lines, across years.

\begin{figure}[ht!]
\captionsetup[subfigure]{justification=centering}
     \centering
     \begin{subfigure}[b]{0.24\linewidth}
         \centering
         \includegraphics[width=\textwidth]{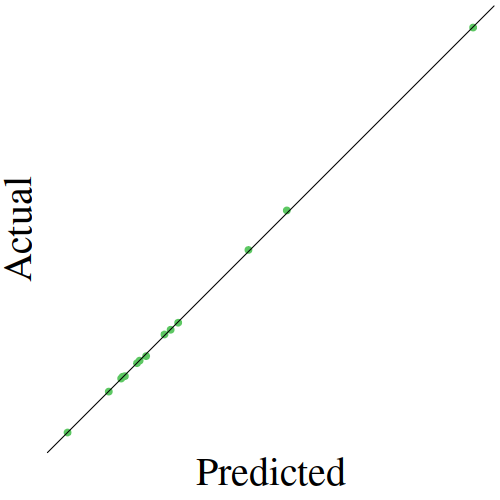}
         \caption{Régions (14)}
         \label{fig:sim_pop13_tvp_reg}
     \end{subfigure}
     \begin{subfigure}[b]{0.24\linewidth}
         \centering
         \includegraphics[width=\textwidth]{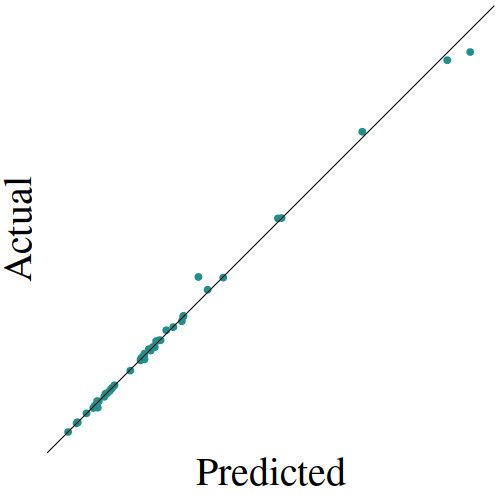}
         \caption{Départments (45)}
         \label{fig:sim_pop13_tvp_dept}
     \end{subfigure}
     \begin{subfigure}[b]{0.24\linewidth}
         \centering
         \includegraphics[width=\textwidth]{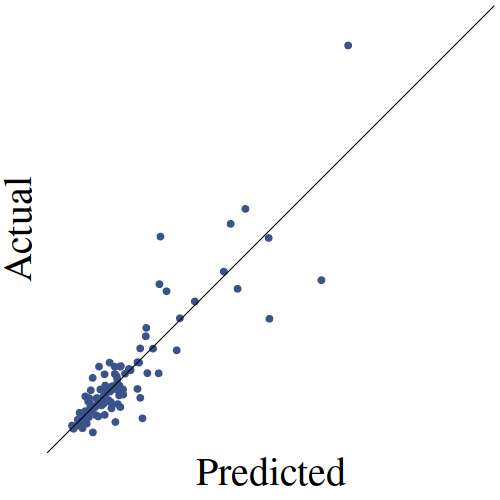}
         \caption{Arrondissements (103)}
         \label{fig:sim_pop13_tvp_arr}
     \end{subfigure}
     \begin{subfigure}[b]{0.24\linewidth}
         \centering
         \includegraphics[width=\textwidth]{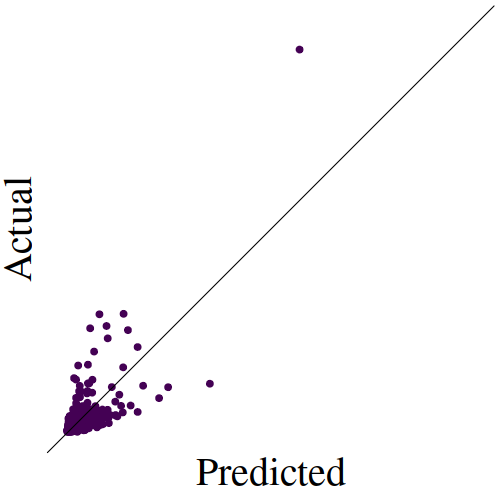}
         \caption{Communes (552)}
         \label{fig:sim_pop13_tvp_com}
     \end{subfigure}
     \hfill
     \begin{subfigure}[b]{0.24\linewidth}
         \centering
         \includegraphics[width=\textwidth]{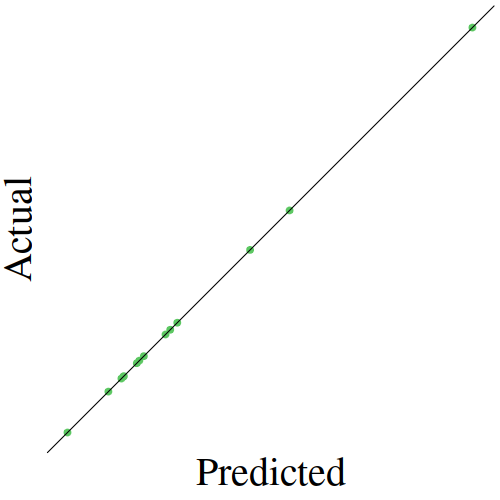}
         \caption{Régions (14)}
         \label{fig:sim_pop13_recalc_tvp_reg}
     \end{subfigure}
     \begin{subfigure}[b]{0.24\linewidth}
         \centering
         \includegraphics[width=\textwidth]{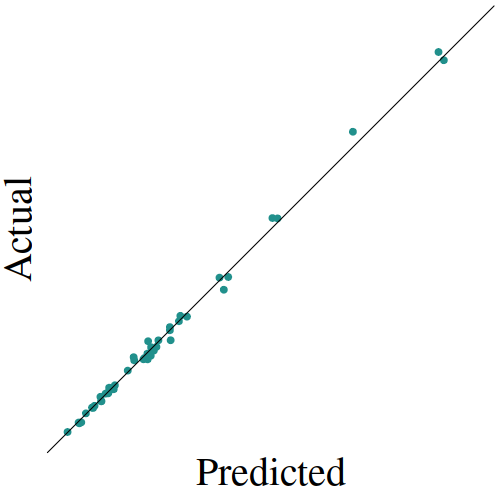}
         \caption{Départments (45)}
         \label{fig:sim_pop13_recalc_tvp_dept}
     \end{subfigure}
     \begin{subfigure}[b]{0.24\linewidth}
         \centering
         \includegraphics[width=\textwidth]{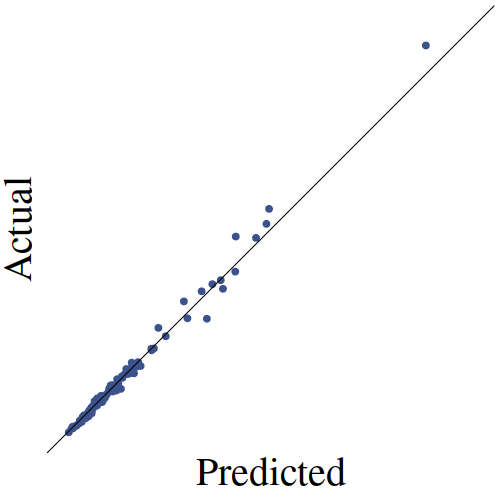}
         \caption{Arrondissements (103)}
         \label{fig:sim_pop13_recalc_tvp_arr}
     \end{subfigure}
     \begin{subfigure}[b]{0.24\linewidth}
         \centering
         \includegraphics[width=\textwidth]{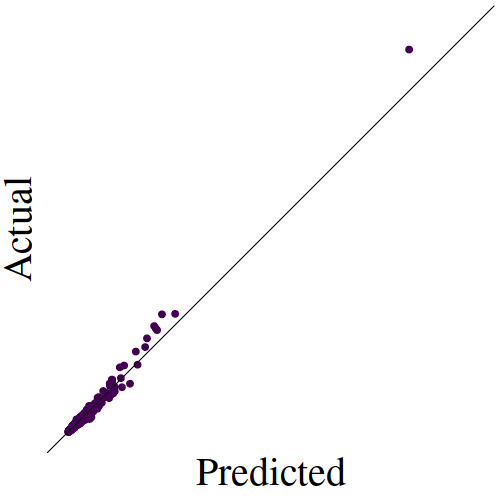}
         \caption{Communes (552)}
         \label{fig:sim_pop13_recalc_tvp_com}
     \end{subfigure}
        \caption{\textbf{Accuracy of satellite imagery-based population estimates across geographies.}\\
        Satellite imagery-derived population estimates (predicted) are compared with RGPHAE 2013 population counts (actual) for original WorldPop estimates trained on département-level (a-d) and recalculated estimates trained on the commune-level (e-h).}
        \label{fig:sim_pop13_tvp}
\end{figure}

WorldPop data are provided in 
tagged image file format (TIFF) with a pixel representing roughly a 100m $\times$ 100m grid square in an open data repository under CC4.0 licence (\cite{WorldPopwww.worldpop.org-SchoolofGeographyandEnvironmentalScienceUniversityofSouthamptonDepartmentofGeographyandGeosciencesUniversityofLouisvilleDepartementdeGeographie2018WorldPopb1}). We aggregate the pixel values via their centroids to the respective administrative areas in Senegal. For discussions on the impact of allocation uncertainty across different area systems we refer to \cite{Koebe2020BetterModelling} and  \cite{Gro2020SwitchingBerlin}. While WorldPop already provides population estimates for Senegal in their open data repository, we opt to slightly modify their approach to align it with our case study. The reasons for this are two-fold: First, the original approach maps RGPHAE 2013 data dasymetrically via a pixel-level weighting layer derived from a hierarchical random forest model trained on départment-level data from the same census and satellite imagery-based covariates from 2013. As Figures \ref{fig:sim_pop13_tvp_reg} to \ref{fig:sim_pop13_tvp_com} show, accuracy of the satellite imagery-derived population estimates steeply decline for small areas, hinting at little spatial robustness of the underlying estimation method. Second, temporal trends in the relationship between population numbers and satellite imagery-based covariates are not accounted for as model training took place at one point in time. For example, while there were on average 1.41 inhabitants per pixel identifying a built-settlement in 2002, this number reduced to 0.88 in 2013. While reasons for that could be manifold (e.g. growth in buildings not used for housing or in per capita housing space) it shows that this could lead to an under- or over-estimation of the population count in 2002 depending on the direction and size of the coefficient.

In contrast, the validation study sets out to update RGPH 2002 tables to 2013, so training needs to take place on 2002 data to make the dynamic population shares comparable to the fixed approach. Moreover, as modelling exercises generally benefit from additional data points, we extend the training phase from originally 45 départements to 103 arrondissements and 552 communes in the validation study and the application, respectively. Comparing Figures  \ref{fig:sim_pop13_tvp_reg} to \ref{fig:sim_pop13_tvp_com} with Figures \ref{fig:sim_pop13_recalc_tvp_reg} to \ref{fig:sim_pop13_recalc_tvp_com} illustrate nicely that this holds true in our case study as well. For further details on the covariates, the preprocessing of the satellite imagery-derived indicators and the hierarchical random forest model used in WorldPop, we refer to \cite{Stevens2015DisaggregatingData} and \cite{Fick2017WorldClimAreas}. Table \ref{tab:worldpop_mod} summarizes the different approaches.

\begin{table}[ht!]
\centering
\begin{tabular}{llll}
\toprule
Approach & Training level & Prediction level & Distribution level \\
\midrule
WorldPop & Département, 2013 & Pixel, 2013 & Département, 2013 \\[1mm]
Validation & Arrondissement, 2002 & Arrondissement, 2013 & Région, 2013 \\[1mm]
Application & Commune, 2013 & Commune, 2013-20 & Région, 2013-20 \\[1mm]
  \hline
\end{tabular}
\caption{\textbf{WorldPop population estimates and their modifications.}\\}
\label{tab:worldpop_mod}
\end{table}

For the validation study, we modify the original approach to fit the time window 2002 to 2013 and to make the results comparable with the proportion method and the application setup. Specifically, we use arrondissement-level data from 2002 (RPGH 2002 and satellite imagery-based covariates from 2002) for fitting the hierarchical random forest model. This way, we make better use of more fine-granular training data and we align the time window for training of the dynamic and the fixed approach. In a next step, we predict the weighting layer for the arrondissement-level of the year 2013 using satellite imagery-based covariates from 2013 and use it for distributing regional-level RGPHAE 2013 population counts to the arrondissements. This allows us to account for both spatial and temporal uncertainty in the evaluation. In addition, initial tests indicate that creating the weighting layer for very small areas, i.e. pixels, introduces much instability also seen in large-area aggregates (see Figure \ref{fig:sim_pop13_tvp}). For example, population estimates for 2013 using the validation setup described above with an arrondissement-level weighting layer correlates highly at $\rho = 0.93$ with the RGPHAE 2013 population counts. In contrast, population estimates derived the same way, but with a pixel-level weighting layer correlate at $\rho = 0.33$.  Therefore, we strike a balance here between spatial disaggregation and robustness by opting for small areas for which we can do both training and testing on. In addition, we compared the accuracy of the population estimates from the hierarchical Bayes approach proposed by \cite{leasure2020national} with the results from the hierarchical random forest, however, without improvements in accuracy.

For the application, we use the approach from the validation study, but apply it on the commune-level for the years 2013 to 2020. Specifically, region-level population projections are mapped to communes using a commune-level weighting layer derived from a hierarchical random forest model trained on commune-level RGPHAE 2013 counts. Not all covariates are available on an annual basis until 2020. In these cases, we consider the latest data available for weighting layer prediction.

\section{Validation}\label{section-validation}

The contribution of this paper is to allow subnational population shares to vary over time (in contrast to fixed shares via the proportion method). For example, 2\% of the population of area A lives in small area A1 during the first census. The proportion method assumes that also in the following years 2\% of the population of area A will live in small area A1, regardless of the actual situation on the ground. Thus, the subnational population shares are considered as fixed. In contrast, we propose to use auxiliary data to capture subnational population shifts during census years. Thus, the subnational population shares can vary over time; we consider them to be dynamic. Arrondissement-level population shares within regions have changed significantly between 2002 and 2013 with growth rates ranging from -91\% to +48\% with the absolute change averaging at 15.7\%. Figure \ref{fig:sim_slope_true} illustrates the true changes in population shares for the ten arrondissements within the région Dakar. Fixed population shares would require the lines to be horizontal, however, this assumption clearly does not hold. The question then is if auxiliary information, in our case satellite imagery, are capable to capture that change more accurately and whether better accuracy translates into the expected improvements for census updating. Concerning the first point, Figure \ref{fig:sim_slope_ppp} gives initial indication for Dakar that this might be the case. 

To examine the overall value added of the proposed methodology we use two census periods in which we update the earlier census (denoted without a time index) to the year of later census (which we denote with a time index $t$) and evaluate our contribution vis-à-vis existing approaches. Specifically, we update the arrondissement-level MPI headcount ratio disaggregated by sex of the head of household extracted from RGPH 2002 to 2013. To do so, we use the DHS 2013 for updating the column margins and RGPH 2002 and population estimates from satellite imagery for updating the row margins with fixed and dynamic population shares, respectively. The updated MPI headcount ratios are evaluated against results from RGPHAE 2013 in terms of relative bias and relative root MSE (RMSE) across arrondissements using $R = 500$ simulation rounds. Therefore, we generate $R = 500$ replicates of the census composition $Y_{aj}$ from RGPH 2002 assuming an independent multinomial distribution in each small area $a$ with parameters $Y_{a}^{\text{Pois},r}$ and $\pi_{aj}$. The population row margin is represented as $Y_{a}^{\text{Pois},r}$ and is obtained using a Poisson distribution $\text{Pois}(\lambda = Y_{a})$ for each small area $a$ independently. The second parameter is defined as $\pi_{aj} =  \frac{Y_{aj}} {Y_{a}}$. This same process is applied to the composition from RGPHAE 2013 ($Y_{aj,t}$) that we use for evaluating the updated MPI headcount ratios. Since the idea is to compare the performance of SPREE using the dynamic approach (Equation \ref{eq-prop-phat}) in contrast to the fixed approach (Equation \ref{eq-prop-pbar}), row margins are generated in two ways: For the fixed approach, we calculate the arrondissement-level population count of 2013 $\hat{Y}_{a,t}^r$ using regional-level population counts $\hat{Y}_{k,t}$ from demographic projections and arrondissement-level population shares $p_a^r$ from each census composition replicate $Y_{aj}^r$ as described in Equation \ref{eq-prop-pbar}. For the dynamic approach, we generate $R = 500$ replicates of the population estimates $\hat{Y}_{a,t}^{\text{Aux}}$ created using the hierarchical random forest approach described in Section \ref{subsubsection-appl-si}. Following Equation \ref{eq-prop-phat}, we use the estimated population shares $\hat{p}_{a,t}^r$ to distribute regional-level population counts $\hat{Y}_{k,t}$ to get arrondissement-level population counts $\hat{Y}_{a,t}^r$ for each simulation round $r$. Then, we obtain $R=500$ replicates of the column margin $\hat{Y}_{j,t}$ by simple resampling from each primary sampling unit of the DHS 2013. Finally, we use $Y_{aj}^r$, $\hat{Y}_{j,t}^r$ and the respective $\hat{Y}_{a,t}^r$ to generate arrondissement-level point estimates $\hat{Y}_{aj,t}^r$ of the MPI headcount ratio in 2013, which we then compare against $Y_{aj,t}^r$ from RGPHAE 2013 using the relative bias and empirical RMSE defined as:



\begin{equation*}
\text{Relative bias} \left(\hat{Y}_{aj,t}\right) = \frac{ \frac{1}{R}  \sum_{r=1}^{R} \left(\hat{Y}^r_{aj,t} - Y^r_{aj,t}\right)}  {\frac{1}{R} \sum_{r=1}^{R} Y^r_{aj,t}},
\end{equation*}
and
\begin{equation*}
    \text{Relative RMSE}\left(\hat{Y}_{aj,t}\right) = \frac{ \sqrt{\frac{1}{R}  \sum_{r=1}^{R} \left(\hat{Y}^r_{aj,t} - Y^r_{aj,t}\right)^2} } {\frac{1}{R} \sum_{r=1}^{R} Y^r_{aj,t}}.
\end{equation*}

Naturally, fixed population shares provide very good accuracy for areas exhibiting little change and poor accuracy for areas exhibiting large fluctuations in the inter-censal population. This pattern indicates that dynamic population margins may be able to improve estimates for areas with large changes in  population shares. Table \ref{tab:sim-pop-diff} supports both hypotheses by grouping the arrondissement-level accuracy measurements into quartiles along the true population change.

\begin{table}[ht]
\centering
\small{
\begin{tabular}{lccc}
\toprule
\multirow{2}{*}{Quartile} &\multicolumn{2}{c}{Relative Bias (in \%)}\\ 
\cmidrule{2-3} 
& dynamic & fixed \\
\hline
Lowest & 17.95 & 28.99 \\ 
2nd & 18.41 & 9.98 \\ 
3rd & 7.57 & -5.44 \\
Highest & 0.68 & -19.60 \\ 
\hline
\end{tabular}
\caption{\textbf{Accuracy of estimated population shares in 2013.}\\
Arrondissements ordered by changes in the true population share between 2002 and 2013. The accuracy is evaluated against the true population shares from RGPHAE 2013.
}
\label{tab:sim-pop-diff}
}
\end{table}

For example, for the 50\% of the 103 arrondissements with the smallest changes in population shares between 2002 and 2013 (2nd and 3rd group), the fixed approach outperforms the dynamic approach. For the arrondissements exhibiting the strongest changes in population share, the dynamic approach provides better accuracy than the fixed approach. A potential reason why both approaches, but especially the dynamic approach, exhibit a tendency to overestimate the changes in the population shares is that the underlying population counts are naturally left-censored at zero. This may induce an upward bias in the population estimates, especially for sparsely populated areas and for population margins with comparatively high volatility.

In total, considering satellite imagery-based population estimates for updating the population margin in SPREE provides more accurate population shares for 42 of the 103 arrondissements in this validation study. Therefore, we see indication that satellite imagery-based population estimates can improve existing approaches used in census updating for deriving small area population margins.

One limitation here is that the knowledge on the question of which small areas show large changes of subnational population shares is usually not known ex-ante. Intuitively, one approach to overcome this could be to use dynamic population shares for areas for which the estimated changes are large. However, volatile satellite imagery-based population margins make this an imperfect selection criteria. Therefore, we opt for a more conservative approach in the application by selecting those small areas for which to use the dynamic population shares that are situated within those 25\% of regions showing the strongest population changes following the regional-level estimates from the official population projections.

\begin{figure}[ht!]
    \centering
    \includegraphics[width=\textwidth]{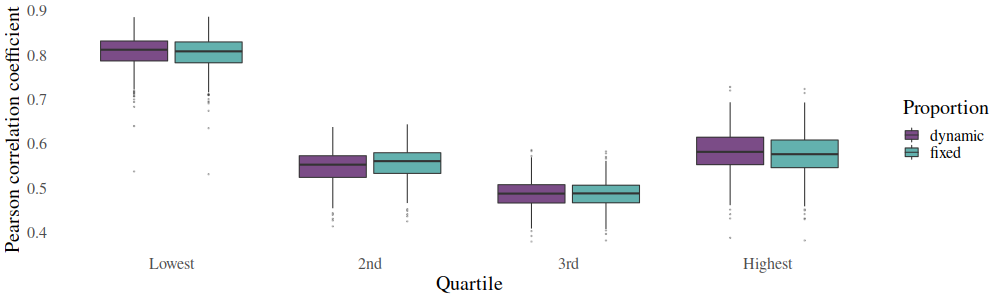}
    \caption{\textbf{Correlation of estimated and true arrondissement-level MPI headcount ratio in 2013.}\\
    Arrondissements are grouped along the quartiles of the changes in true population shares between 2002 and 2013.}
    \label{fig:sim_box_cor}
\end{figure}

Figure \ref{fig:sim_box_cor} and Table \ref{tab:Design_bias_final} show the results of the simulation. The overall correlation of the estimated MPI headcount ratio for 2013 and the true headcount ratio of 2013  as depicted in Figure \ref{fig:sim_box_cor} is moderate for both proportion types. While the Pearson correlation coefficient reflects the overall results from Table \ref{tab:sim-pop-diff}, the differences between the dynamic and fixed approach are less striking. A potential reason is that SPREE builds on multiple inputs (represented by the allocation structure and the association structure), so changes in one of the inputs may not affect the SPREE output to the same extent.

Looking at the relative bias and the relative RMSE in Table \ref{tab:Design_bias_final}, the dynamic approach performs better in most of the cases. The comparatively higher relative bias in the second group can potentially be explained with zero counts in the underlying census composition. As SPREE performs less well in the presence of zero counts and as half of the zero counts in the census composition are located in the second group a deterioration of model performance can be expected. 

\begin{table}[ht!]
\centering
\begin{tabular}{llrrrrrr}
\toprule
\multicolumn{8}{l}{Relative bias (in \%)}\\
\midrule
Quartile & Type & Q2.5 & Q25 & Median & Mean & Q75 & Q97.5 \\
\midrule
Lowest & Dynamic & -29.55 & -11.94 & 1.23 & -1.96 & 7.60 & 22.13 \\ 
& Fixed & -28.92 & -10.39 & 3.94 & -0.24 & 8.64 & 22.87 \\ 
2nd & Dynamic & -14.00 & 4.69 & 15.79 & 15.32 & 27.94 & 43.22 \\ 
& Fixed & -13.57 & 5.05 & 16.87 & 16.02 & 28.55 & 43.38 \\ 
3rd & Dynamic & -24.44 & -0.60 & 2.58 & 6.89 & 16.88 & 36.99 \\ 
& Fixed & -24.15 & 0.19 & 4.40 & 7.74 & 17.18 & 38.17 \\ 
Highest & Dynamic & -30.54 & -6.72 & 1.74 & 3.82 & 18.93 & 34.55 \\ 
& Fixed & -29.83 & -4.28 & 5.50 & 5.58 & 19.37 & 34.83 \\ 
 \toprule
\multicolumn{8}{l}{Relative RMSE (in \%)}\\
\midrule
Quartile & Type & Q2.5 & Q25 & Median & Mean & Q75 & Q97.5 \\
\midrule
Lowest & Dynamic & 7.99 & 13.15 & 17.42 & 18.70 & 23.82 & 32.71 \\ 
& Fixed & 8.44 & 12.76 & 17.42 & 18.23 & 21.96 & 32.12 \\ 
2nd & Dynamic  & 9.12 & 13.81 & 22.51 & 24.09 & 34.36 & 50.96 \\ 
& Fixed & 8.76 & 14.23 & 22.55 & 24.25 & 34.60 & 51.02 \\ 
3rd & Dynamic & 5.91 & 10.02 & 18.36 & 19.99 & 28.46 & 42.52 \\ 
& Fixed & 5.82 & 9.93 & 18.40 & 20.04 & 28.55 & 43.25 \\ 
Highest & Dynamic & 9.06 & 16.03 & 23.16 & 23.59 & 30.25 & 43.14 \\ 
& Fixed & 8.90 & 16.73 & 22.61 & 23.64 & 32.06 & 43.54 \\
  \hline
\end{tabular}
\caption{\textbf{Performance of the 2013 census updates for the MPI headcount ratio in Senegal, across arrondissements.}}
\label{tab:Design_bias_final}
\end{table}


A possible reason why the differences between the dynamic and the fixed approach are less striking in terms of relative RMSE is that the resampling procedure of the respective population margins differs. While the fixed population margin is solely derived from census data, which is resampled following a Poisson and a Multinomial distribution in our setup, the procedure to obtain the dynamic population margin based on satellite imagery considers a simple resampling in order to account for potential non-trivial uncertainty in the pre-processing. This may lead to efficiency losses in the dynamic approach.

In summary, the results of the validation study show that the dynamic approach using satellite imagery from WorldPop as auxiliary information can improve the accuracy of both the population margin (relative bias) and the census updates (in Pearson correlation coefficient), especially for areas with large changes in population shares. For those areas with minor changes in their population shares, the fixed approach is recommended. 

\section{Application}\label{section-application}

In this section we present the results of our case study where we produce updated MPI headcount ratios for all 552 communes in Senegal for the years 2013 to 2020. We apply the semiparametric bootstrap procedure with $B = 100$ explained in Section \ref{subsection-uncertainty} to obtain uncertainty measures for the estimated MPI headcount ratios.

\begin{figure}[ht!]
\begin{minipage}{0.89\textwidth}
\captionsetup[subfigure]{justification=centering}
     \centering
     \begin{subfigure}[b]{0.24\linewidth}
         \centering
         \includegraphics[width=\textwidth]{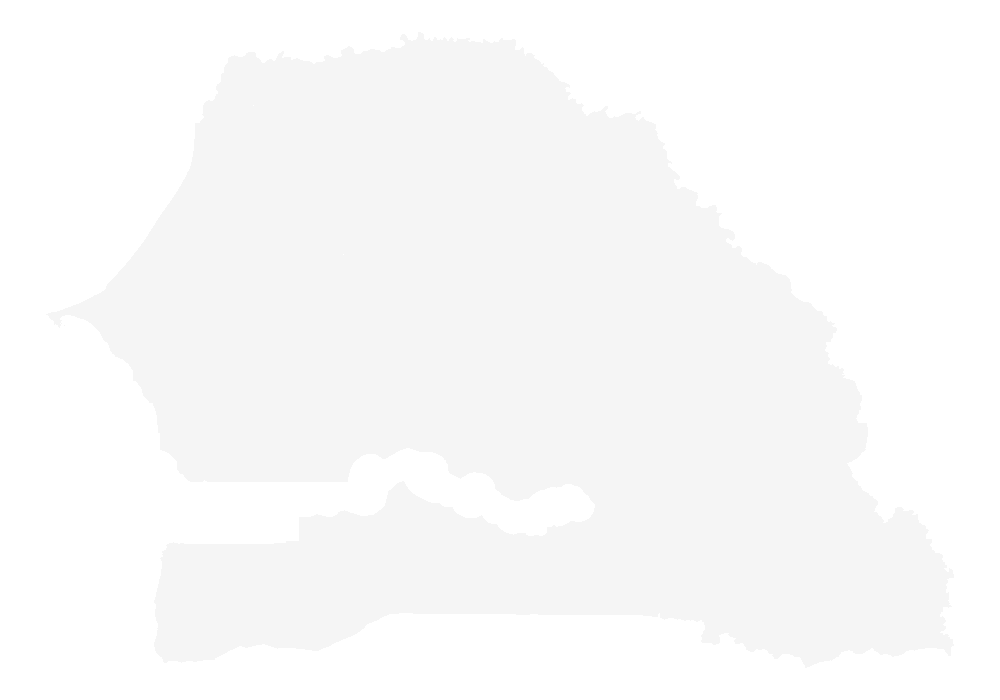}
         \caption{2013}
         \label{fig:appl-mpi-maps-2013}
     \end{subfigure}
     \begin{subfigure}[b]{0.24\linewidth}
         \centering
         \includegraphics[width=\textwidth]{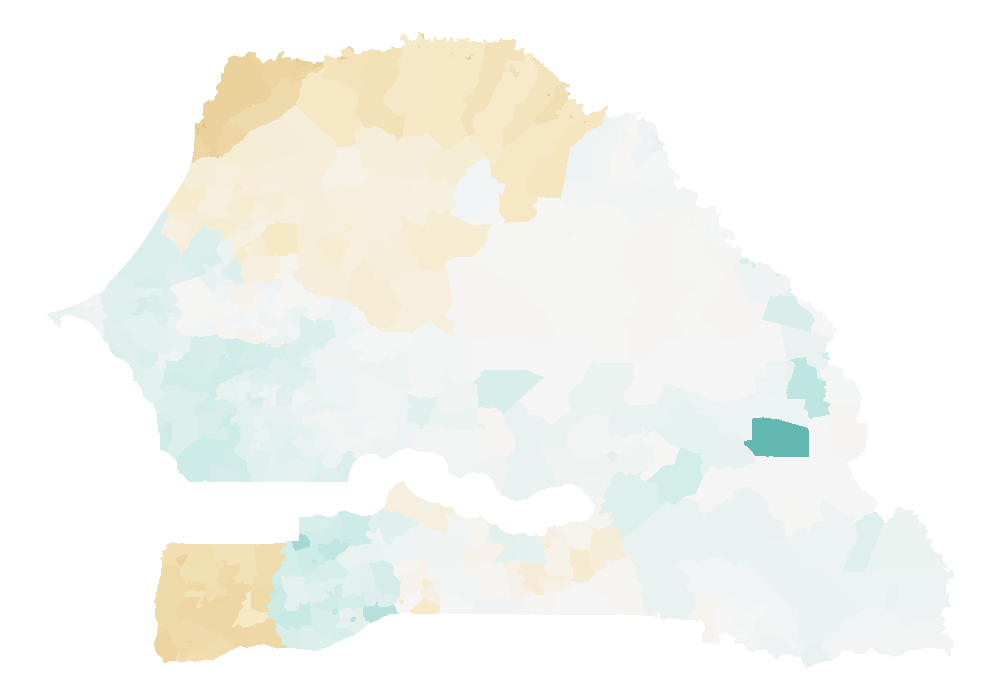}
         \caption{2014}
         \label{fig:appl-mpi-maps-2014}
     \end{subfigure}
     \begin{subfigure}[b]{0.24\linewidth}
         \centering
         \includegraphics[width=\textwidth]{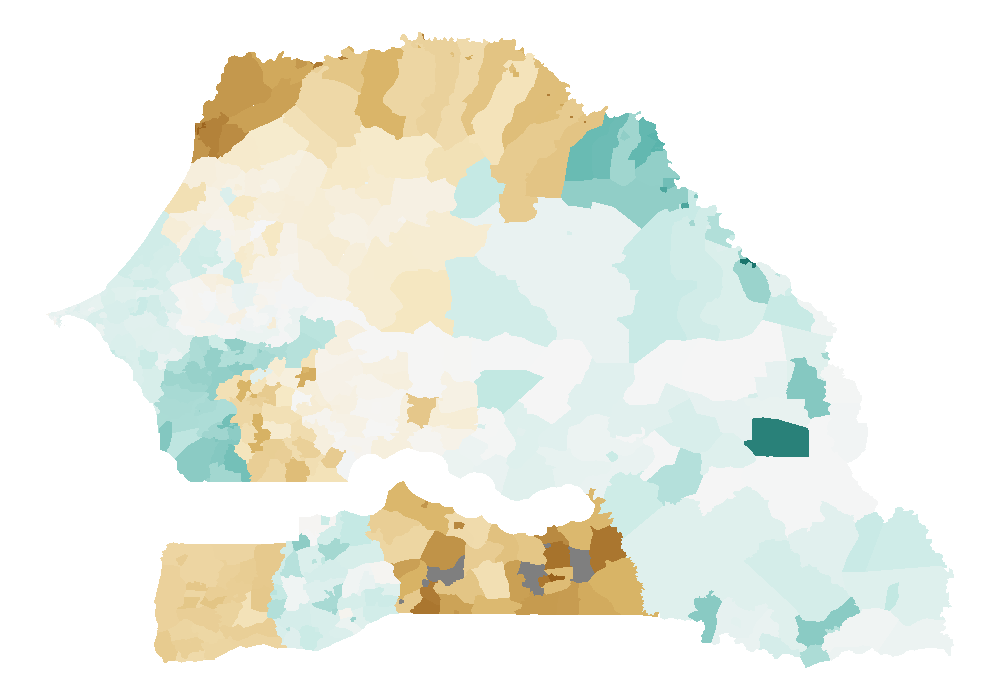}
         \caption{2015}
         \label{fig:appl-mpi-maps-2015}
     \end{subfigure}
     \begin{subfigure}[b]{0.24\linewidth}
         \centering
         \includegraphics[width=\textwidth]{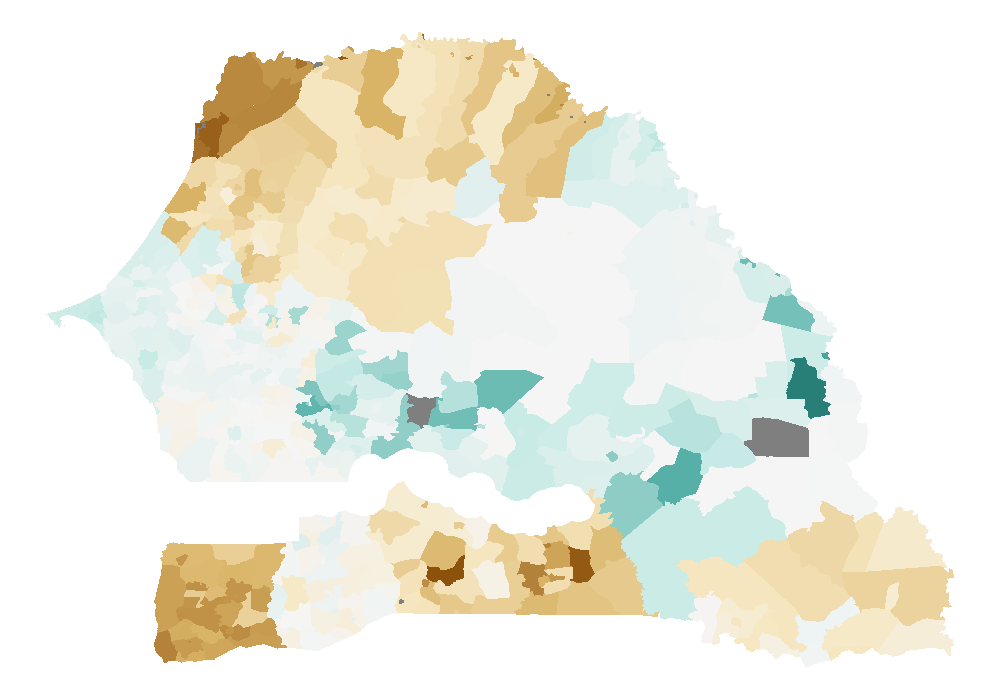}
         \caption{2016}
         \label{fig:appl-mpi-maps-2016}
     \end{subfigure}
     \hfill
     \begin{subfigure}[b]{0.24\linewidth}
         \centering
         \includegraphics[width=\textwidth]{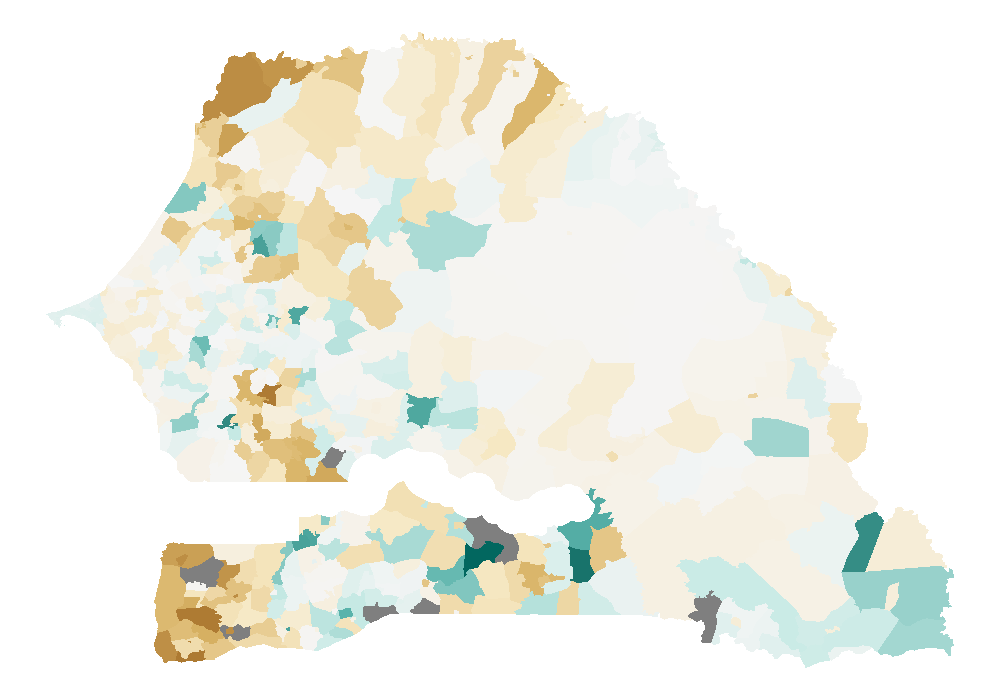}
         \caption{2017}
         \label{fig:appl-mpi-maps-2017}
     \end{subfigure}
     \begin{subfigure}[b]{0.24\linewidth}
         \centering
         \includegraphics[width=\textwidth]{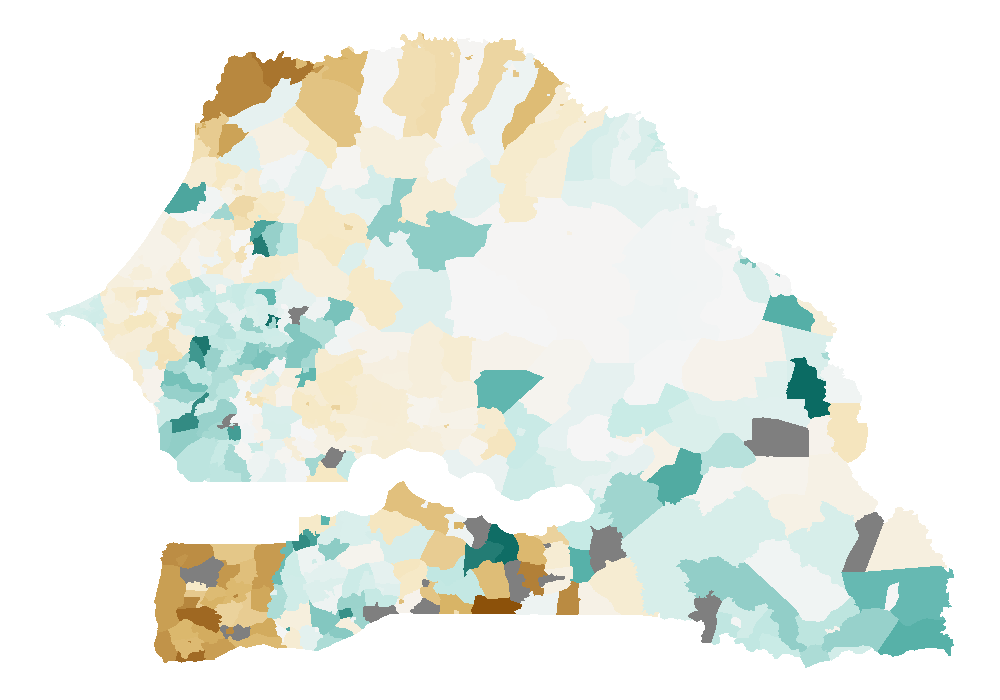}
         \caption{2018}
         \label{fig:appl-mpi-maps-2018}
     \end{subfigure}
     \begin{subfigure}[b]{0.24\linewidth}
         \centering
         \includegraphics[width=\textwidth]{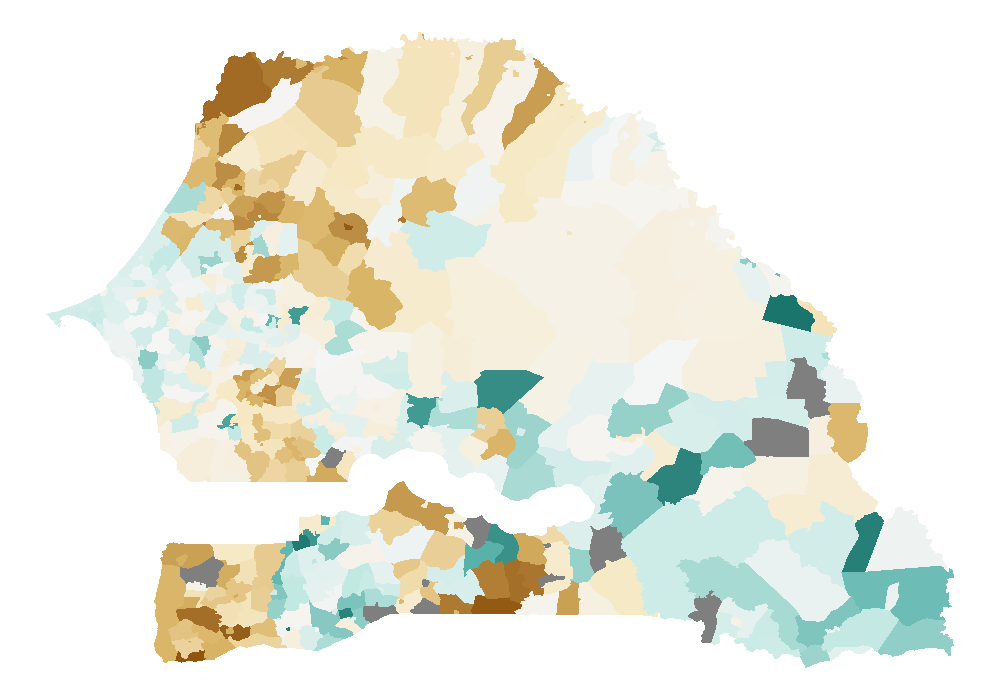}
         \caption{2019}
         \label{fig:appl-mpi-maps-2019}
     \end{subfigure}
     \begin{subfigure}[b]{0.24\linewidth}
         \centering
         \includegraphics[width=\textwidth]{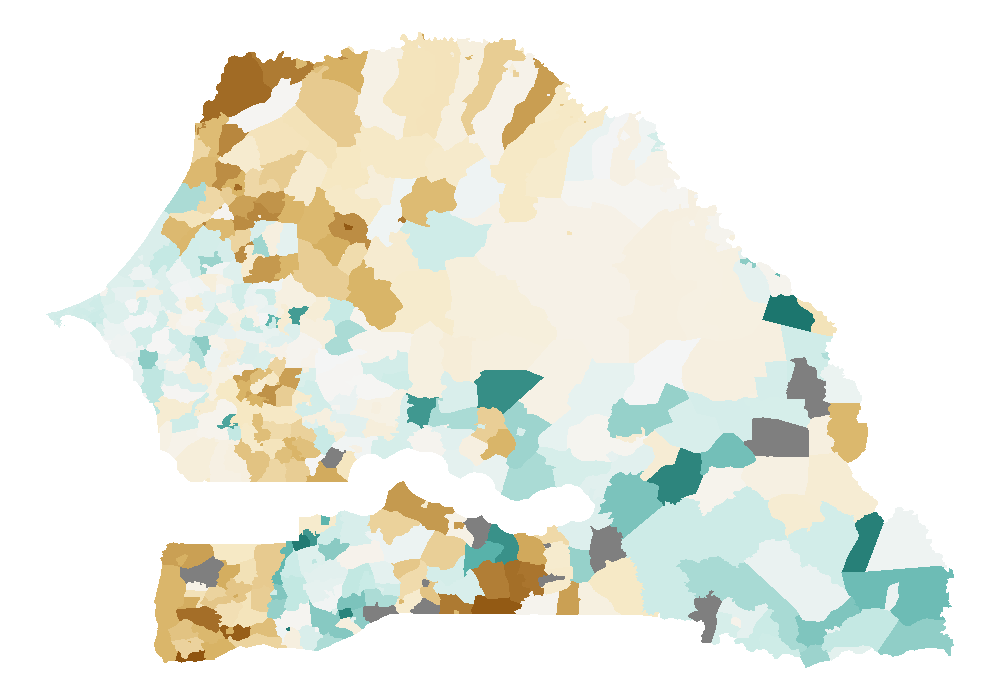}
         \caption{2020}
         \label{fig:appl-mpi-maps-2020}
     \end{subfigure}
     \end{minipage}
    \begin{minipage}{0.1\textwidth}
    \centering
    \includegraphics[width=\textwidth]{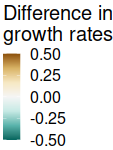}
    \end{minipage}
        \caption{\textbf{Evolution of the updated MPI headcount ratio of individuals living in female-headed households vs. male-headed households.}\\}
        \label{fig:appl-mpi-maps}
\end{figure}

Has the multidimensional headcount ratio of individuals living in female-headed households improved more or less than the one in male-headed households since 2013? As shown in Table \ref{tab:census-h}, individuals living in female-headed households in Senegal are less likely to be multidimensional poor than individuals living in male-headed households. Consequently, it is expected that female-headed households may profit on average less from the social safety net program PNBSF. Figure \ref{fig:appl-mpi-maps} depicts the difference of growth rates in the MPI headcount ratio between individuals living in female-headed households and individuals living in male-headed households using the year 2013 as baseline. The figure and the national-level point estimates of the MPI headcount ratio $H$ in Tables \ref{tab:appl_cv} and \ref{tab:appl_cv_male} confirm that the MPI headcount ratio of female-headed households has improved on average less since 2013 than of male-headed households (represented by the shades of brown in Figure  \ref{fig:appl-mpi-maps}).

Table \ref{tab:appl_cv} shows that the estimated MPI headcount ratio for female-headed households has been constantly decreasing from 2013 to 2020, with a greater reduction between 2016 and 2018. As of 2018, the incidence of multidimensional poverty of this specific population in Senegal has remained stagnant (61\%). The corresponding coefficients of variation (CV) indicate that the updated incidence of multidimensional poverty are acceptable considering traditional thresholds (below 20\%).

\begin{table}[ht!]
\centering
\begin{tabular}{lrrrrrrr}
\toprule
\multirow{2}{*}{Year} & \multicolumn{1}{c}{H} &\multicolumn{6}{c}{Coefficient of variation (in \%)}\\ 
 & (est.) & Q2.5 & Q25 & Median & Mean & Q75 & Q97.5 \\
\midrule
2013 & 0.70 & 0.13 & 5.26 & 9.67 & 12.68 & 14.22 & 29.18 \\ 
  2014 & 0.69 & 0.08 & 3.77 & 7.19 & 7.55 & 10.04 & 21.21 \\ 
  2015 & 0.69 & 0.08 & 4.46 & 8.08 & 9.25 & 12.01 & 20.94 \\ 
  2016 & 0.69 & 0.08 & 4.26 & 8.18 & 9.11 & 12.93 & 21.42 \\ 
  2017 & 0.65 & 0.06 & 3.80 & 6.69 & 6.93 & 9.49 & 16.12 \\ 
  2018 & 0.61 & 0.09 & 6.33 & 10.81 & 10.69 & 14.18 & 24.52 \\ 
  2019 & 0.61 & 0.13 & 6.52 & 10.62 & 11.08 & 13.88 & 28.72 \\ 
  2020 & 0.61 & 0.12 & 6.62 & 10.52 & 10.98 & 13.89 & 27.65 \\  
  \hline
\end{tabular}
\caption{\textbf{Estimated headcount ratio and its coefficient of variation for female-headed households, across communes.}\\}
\label{tab:appl_cv}
\end{table}

In comparison, the MPI headcount ratio for male-headed households (see Table \ref{tab:appl_cv_male}) decreased by 13 percentage points over the same period of time showing considerable smaller CVs than the female-headed households. 
This result is expected since the the sample size of male-headed households is considerably larger (approximately 77.7\% of Senegalese households are headed by men).

\begin{table}[ht!]
\centering
\begin{tabular}{lrrrrrrr}
\toprule
\multirow{2}{*}{Year} & \multicolumn{1}{c}{H} &\multicolumn{6}{c}{Coefficient of variation (in \%)}\\ 
 & (est.) & Q2.5 & Q25 & Median & Mean & Q75 & Q97.5 \\
\midrule
2013 & 0.77 & 0.66 & 2.17 & 3.95 & 4.43 & 6.29 & 9.77 \\ 
  2014 & 0.77 & 0.51 & 1.61 & 2.85 & 3.38 & 4.96 & 7.67 \\ 
  2015 & 0.77 & 0.61 & 2.08 & 3.74 & 4.44 & 6.64 & 9.91 \\ 
  2016 & 0.74 & 0.58 & 2.44 & 4.40 & 4.96 & 7.38 & 10.90 \\ 
  2017 & 0.68 & 0.84 & 2.56 & 3.94 & 4.20 & 5.97 & 8.11 \\ 
  2018 & 0.66 & 0.86 & 3.32 & 5.48 & 6.03 & 8.93 & 12.14 \\ 
  2019 & 0.64 & 1.42 & 4.06 & 6.13 & 6.45 & 8.86 & 12.32 \\ 
  2020 & 0.64 & 1.45 & 4.06 & 6.11 & 6.50 & 8.99 & 12.48 \\ 
  \hline
\end{tabular}
\caption{\textbf{Estimated headcount ratio and its coefficient of variation for male-headed households, across communes.}\\}
\label{tab:appl_cv_male}
\end{table}

\begin{figure}[ht!]
\captionsetup[subfigure]{justification=centering}
    \centering
    \begin{subfigure}[b]{1\linewidth}
         \centering
         \includegraphics[width=\textwidth]{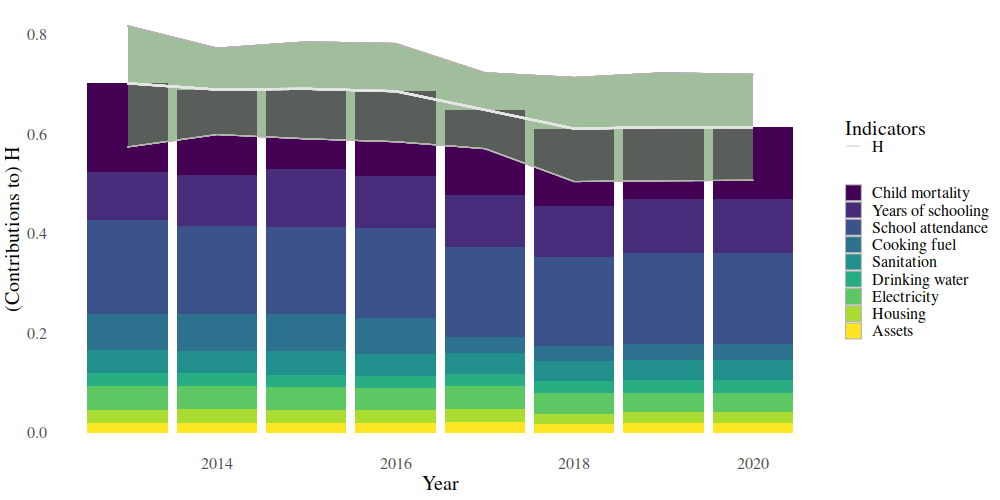}
         \caption{Senegal}
         \label{fig:appl-com-contribution-bar-sen}
     \end{subfigure}
     \hfill
     \begin{subfigure}[b]{\linewidth}
         \centering
         \includegraphics[width=\textwidth]{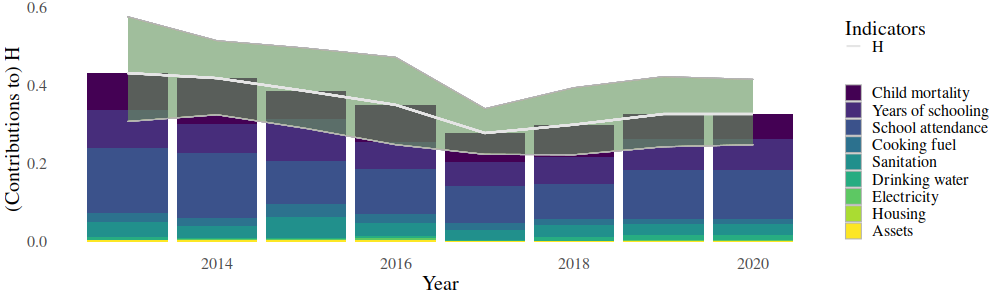}
         \caption{Yoff, Dakar, Senegal}
         \label{fig:appl-com-contribution-bar-yoff}
     \end{subfigure}
    \caption{\textbf{Indicator-specific contributions to the uncensored MPI headcount ratio for individuals living in female-headed households in Senegal and in the commune Yoff located in the region Dakar, specifically.}\\}
    \label{fig:appl-com-contribution-bar}
\end{figure}

Looking at the indicator-specific contributions to the MPI headcount ratio at a national-level, decreases in multidimensional poverty are mainly due to improved living conditions (see Figure \ref{fig:appl-com-contribution-bar-sen}). Notably, deprivation in the indicator on `years of schooling' has increased for individuals living in female-headed households with `school attendance' remaining a main driver of multidimensional poverty. 

In order to drill down to local realisations of national-level trends, we apply SPREE following the hybrid approach laid out in Section \ref{section-validation}, i.e. apply the dynamic approach in those areas with bigger changes in their population shares and the fixed approach for those with less changes. Specifically, the regions Kaffrine, Matam and Tambacounda show the highest population change. Thus, we apply the dynamic margins for the 105 communes located in these regions. Figure \ref{fig:appl-com-contribution-bar-yoff} illustrates the development of indicator-specific contributions to the commune-level MPI headcount ratio of the commune `Yoff' in the region Dakar from 2013 to 2020 as an example. While in Yoff living conditions as drivers of multidimensional poverty steadily declined over the years similar to the national trend, also school attendance has improved, however, still remaining the biggest driver of commune-level MPI poverty.

In this setting neither direct nor indirect evaluation of commune-level census updates is possible due to the lack of appropriate commune-level evaluation data after the census year 2013 on the one hand and due to the fact that SPREE results reproduce region-level survey results by design on the other.

\section{Conclusion}\label{section-conclusion}
This study has demonstrated in a case study how auxiliary information such as satellite imagery can be used to potentially improve the availability of relevant key statistics for small areas between census years by strengthening the population margins especially in regions exhibiting strong population growth. This is particularly important for local-level policy-making as censuses and surveys fail to provide reliable and up-to-date data on small areas. This study has laid out a path to fill this gap by proposing a unified framework to combine established census updating techniques with recent approaches that add spatial granularity to existing data using novel data sources.
We discussed both power and limitations of the proposal by providing annual MPI headcount ratio updates including indicator-specific contributions for the years 2013 to 2020 in the 552 communes of Senegal.

First and foremost, the value added vis-à-vis existing census updating methodologies is largely determined by the quality of the auxiliary data. The simulation showed that conveniently preprocessed remote sensing products such as the population grids including the corresponding covariates currently still lack the spatial robustness needed to add value to existing approaches. While this challenge is likely to be solved in the mid-term by the rapidly evolving market of remote sensing products, it may require considerable effort in the short-term to overcome this quality gap.

Second, rigorously addressing the uncertainty involved in the estimation process demands further research as for many promising auxiliary data sources the data generating process is beyond the control of a national statistical office. In addition, census and survey inputs sometimes require imputation to address high levels of non-response. While we took a first step in this paper to capture parts of the uncertainty by resampling the inputs, it would be interesting to incorporate uncertainty in the explanatory variables and effects from (multiple) imputation more formally.

Third, as we showcased in the case study, slightly varying definitions of the indicator of interest across data sources may induce additional uncertainty into the updating process, so harmonisation along the lines of \cite{Isidro2016ExtendedPovertyb} is clearly advised. This approach could also be used to obtain a complete MPI when some of the components are not present in the census, which is the case in the setup of this paper. 

Fourth, under the SPREE approach, census, survey and auxiliary data must have the same structure in terms of the disaggregation dimensions. However, considering that a census is conducted in long intervals, accounting for re-structurings such as land reforms that alter the administrative delineation, takes relevance. High-resolution population grids allow for a re-allocation of the population across different geographies under certain assumptions, however, a more general approach is desirable.

Fifth, the applicability of the SPREE could also increase significantly by allowing for non-categorical indicators and non-linear relationships. Here, copulas as used in \cite{Rocher2019EstimatingModels} may provide a powerful vehicle to generalize SPREE further.

\section*{Conflict of interest}
The authors have declared that no competing interests exist.

\bibliography{refs}

\end{document}